\documentclass[aps,reprint,superscriptaddress]{revtex4-2}
\usepackage[T1]{fontenc}
\usepackage{physics}
\usepackage{amsmath,amsfonts,bbm, graphicx, color,natbib}
\usepackage{amssymb}
\usepackage{appendix}
\usepackage{relsize}
\usepackage{xcolor}
\usepackage{scalerel}
\usepackage{tikz}
\usepackage{pdfpages}
\usetikzlibrary{svg.path}
\usepackage[normalem]{ulem}

\definecolor{orcidlogocol}{HTML}{A6CE39}
\tikzset{
	orcidlogo/.pic={
		\fill[orcidlogocol] svg{M256,128c0,70.7-57.3,128-128,128C57.3,256,0,198.7,0,128C0,57.3,57.3,0,128,0C198.7,0,256,57.3,256,128z};
		\fill[white] svg{M86.3,186.2H70.9V79.1h15.4v48.4V186.2z}
		svg{M108.9,79.1h41.6c39.6,0,57,28.3,57,53.6c0,27.5-21.5,53.6-56.8,53.6h-41.8V79.1z M124.3,172.4h24.5c34.9,0,42.9-26.5,42.9-39.7c0-21.5-13.7-39.7-43.7-39.7h-23.7V172.4z}
		svg{M88.7,56.8c0,5.5-4.5,10.1-10.1,10.1c-5.6,0-10.1-4.6-10.1-10.1c0-5.6,4.5-10.1,10.1-10.1C84.2,46.7,88.7,51.3,88.7,56.8z};
	}
}

\newcommand\orcid[1]{\href{https://orcid.org/#1}{\mbox{\scalerel*{
				\begin{tikzpicture}[yscale=-1,transform shape]
				\pic{orcidlogo};
				\end{tikzpicture}
			}{|}}}}
\usepackage[colorlinks]{hyperref}

\hypersetup{
	bookmarksnumbered,
	pdfstartview={FitH},
	citecolor={darkgreen},
	linkcolor={darkblue},
	urlcolor={darkblue},
	pdfpagemode={UseOutlines}}
\definecolor{darkgreen}{RGB}{20,100,20}
\definecolor{darkblue}{RGB}{0,0,130}
\definecolor{darkred}{rgb}{.8,0,0}

\usepackage{soul}

\makeatletter
\AtBeginDocument{\let\LS@rot\@undefined}
\makeatother

\begin{document}
	
	\title{Repulsive dynamics of strongly attractive one-dimensional quantum gases}
	
	\author{Maciej \L{}ebek\orcid{0000-0003-4858-2460}}
	\email{mlebek@cft.edu.pl}
	\affiliation{Center for Theoretical Physics, Polish Academy of Sciences, Aleja Lotnik\'ow 32/46, 02-668 Warsaw, Poland}
	\affiliation{Institute of Theoretical Physics, University of Warsaw, Pasteura 5, 02-093 Warsaw, Poland}
	
	\author{Andrzej Syrwid\orcid{0000-0002-0973-4380}}
	\email{syrwid@kth.se}
	\affiliation{Department of Physics, KTH Royal Institute of Technology, SE-106 91 Stockholm, Sweden}
	
	\author{Piotr T. Grochowski\orcid{0000-0002-9654-4824}}
	\email{piotr@cft.edu.pl}
	\affiliation{Center for Theoretical Physics, Polish Academy of Sciences, Aleja Lotnik\'ow 32/46, 02-668 Warsaw, Poland}
	
	\author{Kazimierz Rz\k{a}\.zewski\orcid{0000-0002-6082-3565}}
	\email{kazik@cft.edu.pl}
	\affiliation{Center for Theoretical Physics, Polish Academy of Sciences, Aleja Lotnik\'ow 32/46, 02-668 Warsaw, Poland}
	\date{\today}
	
	\begin{abstract}
		We analyze the dynamics of one-dimensional quantum gases with strongly attractive contact interactions.
		We specify a class of initial states for which attractive forces effectively act as strongly repulsive ones during the time evolution.
		Our findings extend the theoretical results on the super-Tonks-Girardeau gas to a highly nonequilibrium dynamics.
		The novel mechanism is illustrated on the prototypical problem of the domain stability in a two-component Fermi gas.
		We also discuss finite-range interactions and analyze universality of the presented results.
		Moreover, we use our conclusions to argue for the existence of metastable quantum droplets in the regime of strongly attractive contact and attractive dipolar interactions.
	\end{abstract}

	\maketitle
	\textbf{Introduction.}---Modern experimental techniques allow for realization of one-dimensional (1D) many-body systems with well controlled interparticle interactions.
	Both the infinitely repulsive Tonks-Girardeau (TG) \cite{Girardeau1960,Kinoshita2004,Paredes2004} and strongly attractive regime can be investigated by means of Feshbach~\cite{Chin2010} or confinement-induced resonances~\cite{Haller2010}.
	Importantly, there are several extraordinary phenomena such as fermionization of strongly repulsive bosons that occur only in 1D~\cite{Giamarchi2007,Zurn2012}.
	Moreover, in 1D there are Bethe-Ansatz-solvable models \cite{Oelkers2006,Gaudin2012,Takahashi1999} such as the bosonic Lieb-Liniger (LL)~\cite{Lieb1963,Lieb1963a} and fermionic Yang-Gaudin (YG)~\cite{Yang1967,Guan2013} ones where analytical solutions provide deep insight into highly correlated many-body physics, far beyond the applicability of mean-field methods. 
	
	Here, we investigate the case of strongly attractive contact interactions, $V(x)=g\delta(x)$ with $g<0$, where significant interparticle correlations and formation of deep bound states are expected~\cite{Thacker1981,Castin2001,Syrwid2021,Calabrese2007,Piroli2016,Piroli2016a,Sykes2007}.
	Interestingly, strongly attractive forces may also support gaslike states that are stable against the collapse~\cite{Astrakharchik2005,Haller2009,Kao2021,Chen2010}.
	Namely, it turns out that the ground state of infinitely repulsive bosons does not collapse when quenching to a very strong interatomic attraction.
	Instead, it stays in a metastable state called super-Tonks-Girardeau (sTG) gas exhibiting anticorrelations characteristic for the TG gas.
	This phenomenon was explained basing on exact solutions of the LL model in~\cite{Batchelor2005a}, where the authors identified a highly excited eigenstate of the strongly attractive system having a large overlap with the initial TG state.
	Similar states were also found in other models~\cite{Chen2010a,Guan2010,Girardeau2010,Astrakharchik2008,Girardeau2012,Girardeau2011}.  
	
	In the strongly attractive regime, spectra of all these systems share a common trait---in addition to deeply bound states describing pairs or clusters of particles, they contain states very close to certain eigenstates of the infinitely repulsive system.
	In~\cite{Durr2009} the spectrum of strongly attractive LL model was identified, revealing that in this regime eigenstates are either bound or close to the TG gas eigenstates.
	Similar observation was made for confined few-body systems~\cite{Gharashi2013,Lindgren2014,Bugnion2013}. 
	Finally, it was pointed out in \cite{Volosniev2014} that such a form of the spectrum is universal both for fermions and bosons, independently of a confining potential.

	In this Letter we explore implications of such a structure of many-body eigenstates for the system dynamics.
	We make use of the similarity between some eigenstates in the $g \to -\infty$ regime to eigenstates of a strongly repulsive system, finding dramatic consequences for the dynamical system behavior.
	Strictly speaking, we show that for an arbitrary trapping potential, there exists a class of initial states that in the regime of strong attraction evolve "repulsively", i.e., in the same way as under the Hamiltonian with strongly repulsive interactions. 
	Such initial states are characterized by vanishing probability density of finding two interacting particles at the same position.
	In the course of time evolution, the particles cannot form bound states despite strong attraction and act as if the interactions were strongly repulsive.
	
\begin{figure}[t!]
		\includegraphics[width=\linewidth]{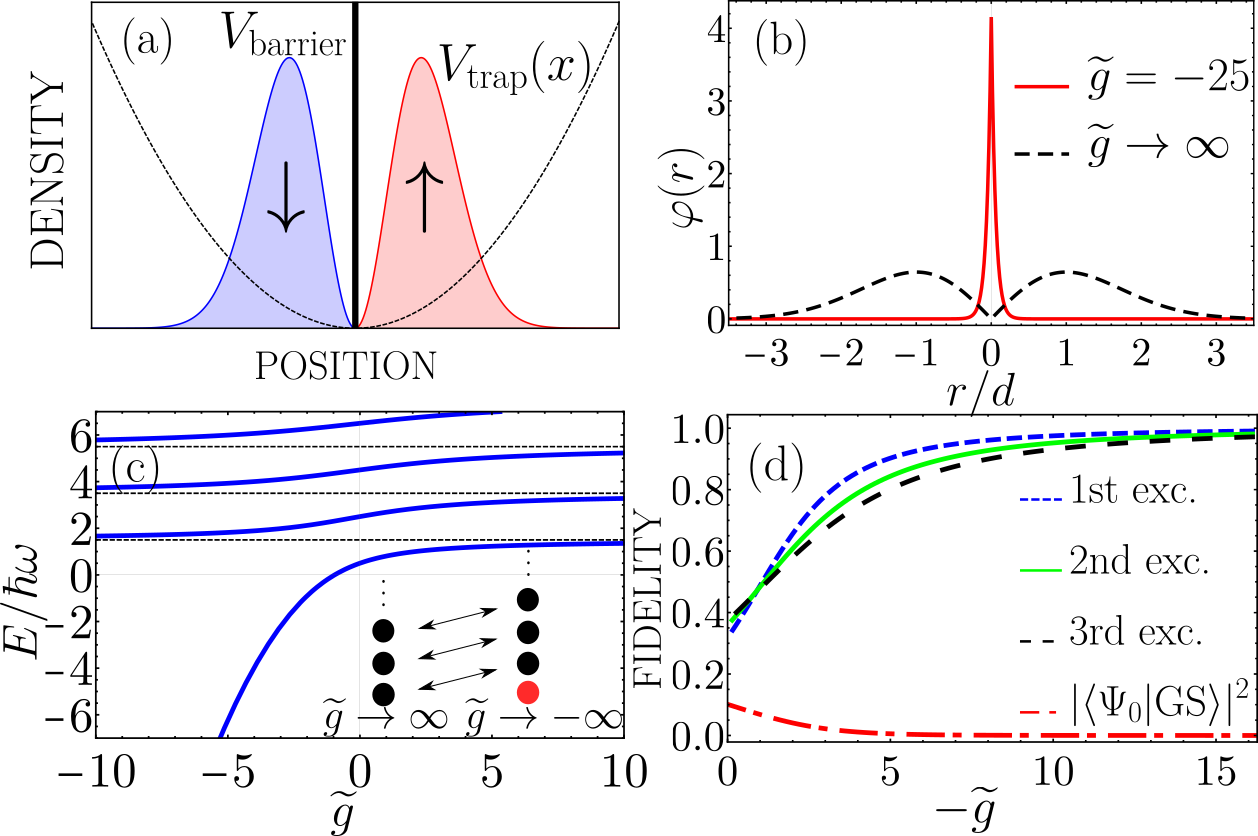}
		\caption{Panel (a): Initial state of the domain wall stability problem in a two-body setup.
			Two strongly interacting fermions of different spins confined in a harmonic trap are separated by a strong potential barrier $V_\text{barrier}$.
			Panel (b) shows ground states of the relative Hamiltonian in the regimes of  infinite repulsion (dashed black line) and strong attraction (red line).
			When decreasing $\widetilde g<0$, the ground state becomes more and more localized around $r=0$.
			Panel (c): energies of symmetric eigenstates of the relative motion versus $\widetilde g$.
			A relationship between spectra for $\widetilde g \to  \pm \infty$ is illustrated in the pictorial inset, where the black dots represent TG states and the red dot represents the bound state.
			Panel (d): influence of $\widetilde g$ on fidelities between first three excited symmetric eigenstates and their TG counterparts.
			The red dash-dotted curve represents the transition probability between the ground state $\left|\text{GS} \right>$  and the initial state $\left| \Psi_0\right>$.
		}\vspace{-0.2cm}
		\label{fig1}
	\end{figure}

	The well-known stability of the sTG phase after the interaction quench from the TG ground state is just one of the consequences of the considered strongly interacting system properties.
	Therefore, our findings may be viewed as an important extension of the sTG behavior to nonequilibrium dynamics.
	In particular, initial states leading to the repulsive evolution of attractive gases can be prepared with geometric quenches  where the sign flip of the interactions is not necessary.
	The surprising mechanism described here is demonstrated on the domain wall stability problem in a harmonic trap~\cite{Massignan2013,Trappe2016,Grochowski2017a,Koutentakis2019,Koutentakis2020,Ryszkiewicz2020,Karpiuk2020,Grochowski2020a,Syrwid2021a,Trappe2021,Valtolina2016,Amico2018,Scazza2020}, a scenario motivated by experiment~\cite{Valtolina2016}, see Fig.~\ref{fig1}(a).
	We show that initially separated domains are stabilized by the strong attraction over many trap periods.
	We also briefly analyze the case of nonlocal interactions.
	Furthermore, insights gained here suggest that bosonic dipolar quantum droplets with strongly repulsive contact interactions~\cite{Odziejewski2020} may survive the quench to the strong contact attraction.
	
	
	\textbf{Domain stability problem.}---First, we consider two distinguishable fermions of mass $m$ confined in a harmonic trap with frequency $\omega$. 
	The corresponding Hamiltonian $H=-\hbar^2/2m( \partial_x^2+\partial_y^2) +\frac{1}{2}m\omega^2(x^2+y^2)+V(x-y)$ is easily separable into center of mass (CM), $R=(x+y)/\sqrt{2}$, and relative, $r=(x-y)/\sqrt{2}$, coordinates and can be solved analytically~\cite{Busch1998}. 
	In general, the eigenstates take the form $\Psi_{n,\mu}(x,y)=\phi_n(R)\varphi_\mu(r)$, where $\phi_{n=0,1,2,\ldots}(R)$ denote the harmonic oscillator eigenstates, while $\varphi_\mu(r)$ can be divided into symmetric and antisymmetric ones (for details see~\cite{Busch1998}).
	The latter are not affected by the contact interactions  due to the vanishing wave function at $r=0$.
	
	We analyze the strongly interacting limits, $\widetilde g = g/\hbar \omega d \to \pm \infty$, where $d=\sqrt{\hbar /m \omega}$, focusing on the relative motion only. 
	While panel (b) of Fig.~\ref{fig1} illustrates relative part of both TG ($\widetilde g \to \infty$) and bound ($\widetilde g<0$) ground states, panel (c) shows eigenenergies corresponding to the symmetric states.
	Note that the spectrum for $\widetilde g\to -\infty$ is similar to the one obtained for $\widetilde g \to \infty$ except the fact that it is supplemented by a deeply bound state with diverging energy.
	Indeed, the ground state in the TG limit corresponds to the first excited symmetric state of strongly attractive system~\cite{Girardeau2011}, as can be verified from their fidelity ($|\cdot |^2$ of the overlap) tending to 1 as $\widetilde g \to -\infty$.
	Similarly, also higher symmetric eigenstates tend to their TG counterparts as $\widetilde g$ becomes more and more negative, see Fig.~\ref{fig1}(d).
	
	Let us now study dynamics of two opposite-spin fermions separated at $t=0$ by a thin and high potential barrier, see Fig.~\ref{fig1}(a). 
	The corresponding initial state reads $\Psi_0(x,y)=2 \phi_1(x)\phi_1(y) \theta(x)\theta(-y)$, where $\theta(x)$ denotes Heaviside step function.
	After the barrier removal, atoms start to push into each other due to the presence of the trap. 
	While, not surprisingly, strong repulsion prevents atoms from mixing, it is striking that the similar system behavior is also observed in the strongly attractive regime.
	\begin{figure}[t!]
		\includegraphics[width=\linewidth]{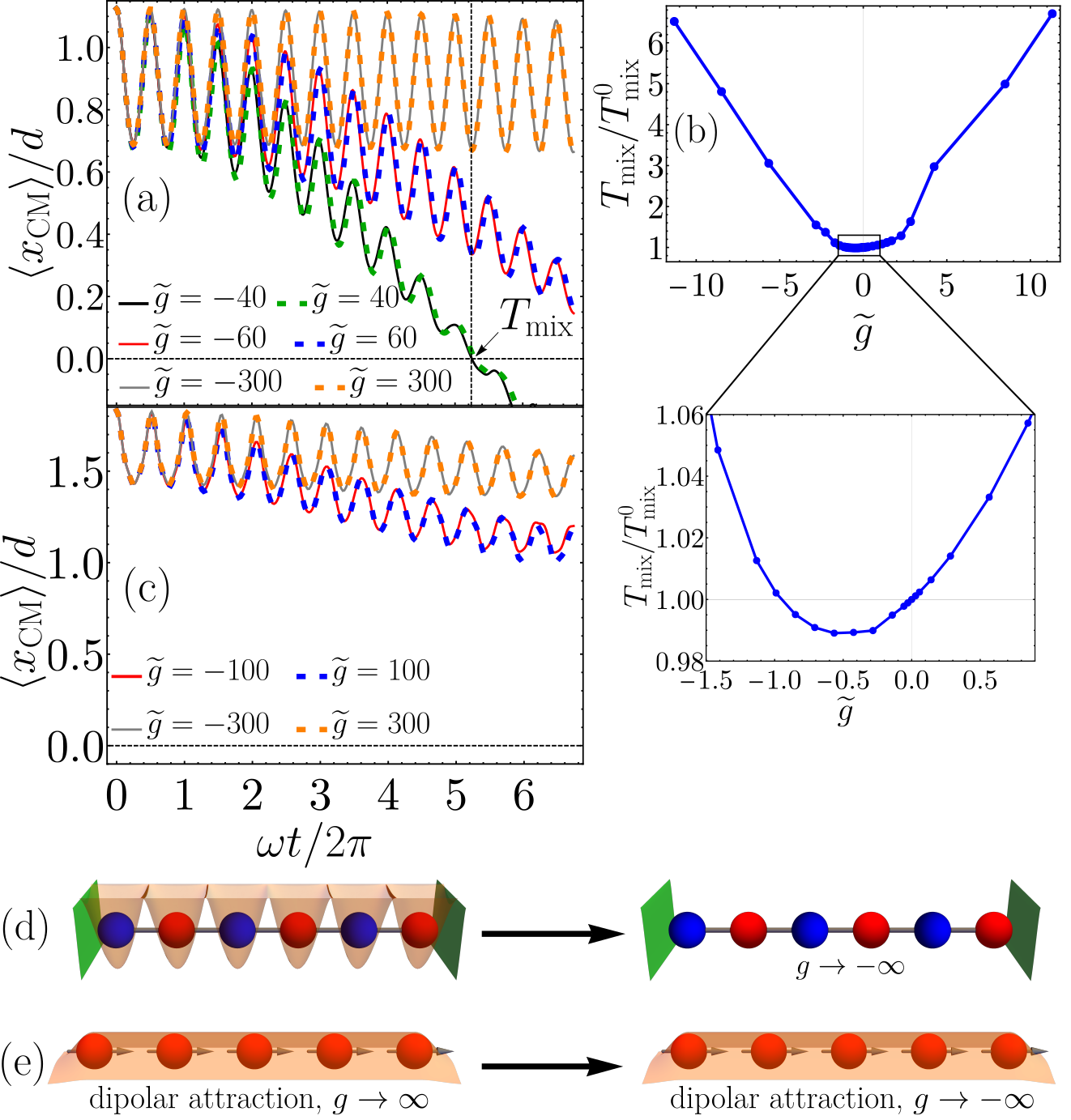}
		\caption{Panel (a): Center of mass position $ \langle x_\text{CM} \rangle$ as a function of time in the two-body problem.
			Both strong repulsion and attraction prevent from mixing.
			Panel (b) presents the two-body mixing time $T_\text{mix}$ versus $\widetilde g$, where in the noninteracting case $T^0_\text{mix}=\pi/2\omega$.
			Panel (c): Temporal behavior of $ \langle x_\text{CM} \rangle$ in the domain stability problem involving six fermions ($N_{\uparrow,\downarrow}=3$).
			As in the two-body case, the domain is stabilized by both strong contact repulsion and attraction.
			Panel (d) illustrates a confinement quench in a two-component Fermi chain, prepared with a deep optical lattice.
			After the lattice removal  the particle order persists despite strong attraction.
			Panel (e): stability of a dipolar quantum droplet, i.e., interaction
			quench from $g \to \infty$ to $g\to -\infty$ does not lead to the droplet collapse.
		}\vspace{-0.2cm}
		\label{fig2}
	\end{figure}
	
	In Fig.~\ref{fig2}(a) we present  the center of mass position $\langle x_\text{CM}\rangle$ calculated for one of the atoms \footnote{Due to the symmetry, the CM position of the second one is obtained via sign change.} as a function of time.
	In the noninteracting case, $ \langle x_\text{CM} \rangle $ oscillates with the trap frequency $\omega$.
	To quantify the effect of interactions on the system dynamics, we introduce the mixing time $T_\text{mix}$ corresponding to the first instance when the atom passes the trap center, i.e., $\langle x_\text{CM} \rangle=0$.
	Fig.~\ref{fig2}(b) shows how $T_\text{mix}$ depends on $\widetilde g$.
	Note that for the weak and moderate attraction, $T_\text{mix}$ slightly decreases reaching a minimum at $\widetilde g \approx -0.5$, which agrees with a basic intuition that interparticle attraction should help the atoms to mix with each other.
	It is no longer true when we go to sufficiently strong interactions, where $T_\text{mix}$ gradually grows.
	For $\widetilde g \to \pm \infty$, the mixing time becomes infinitely long indicating stabilization in the system.
	In order to understand why the interaction $\widetilde g \to -\infty$ yields the same evolution as the limit of infinite repulsion, let us write the explicit form of the time-evolved wave function
	\begin{equation}
	\label{eq::dyn}
	\Psi(x,y;t)=\sum_{n,\mu} e^{-\frac{i}{\hbar}E_{n,\mu}t}C_{n,\mu}\Psi_{n,\mu}(x,y),
	\end{equation}
	where $E_{n,\mu}$ denote energies of eigenstates $\Psi_{n,\mu}(x,y)$ and $C_{n,\mu}=\big< \Psi_{n,\mu}\big| \Psi_0\big>$.
	Note that while our initial state $\Psi_0\big|_{x=y}=0$, all boundlike states are strongly localized at $x=y$ and almost zero elsewhere.
	Therefore, all the corresponding $C_{n,\mu}$ are very small and these bound states effectively do not contribute to the dynamics.
	In Fig.~\ref{fig1}(d) we present the decay of the fidelity between ground state (characterized by a bound state in a relative coordinate) and the initial state $\Psi_0$ with growing $-\widetilde g$.
	Consequently, the dynamics happens mostly in the subspace of TG-like states and therefore closely follows the dynamics of the TG system (becoming identical for $\widetilde g \to -\infty$).
	We emphasize here that the zero value of the initial wave function for $x=y$, together with the special structure of the spectrum, are the key ingredients of the repulsive dynamics. 
	Note, however, that the former requirement is not necessary in the strict limiting cases of $\widetilde g \to \pm \infty$.
	Since the positive energy eigenstates span the whole Hilbert space for $\widetilde g \to \infty$ and each of them coincides with some eigenstate of $\widetilde g \to -\infty$, the bound states identified for finite and negative $\widetilde g$ do not contribute in the limit of the infinite attraction.
	As such, for the infinite interaction, dynamics is always "repulsive" as the overlap between any square-integrable initial wave function and a bound state tends to zero.
	Such a conclusion can also be reached for many-body contact-interacting systems in which eigenstates coincide in the $ g \to \pm \infty$ limits.
	
	A crucial question concerns the presence of the mechanism also in larger systems, with initial state analogous to Fig.~\ref{fig1}(a).
	As the problem lacks an exact solution, we turn to numerical methods employing a standard density matrix renormalization group (DMRG) technique~\cite{White1992,White1993,Schollwock2005,Schollwock2011,Orus2014} and a recently improved algorithm for time evolution where one-site time-dependent variational principle (TDVP) scheme~\cite{Kramer2008,Haegeman2011,Koffel2012,Haegeman2016} is combined with a global basis expansion~\cite{tdvp2020}.
	To perform both DMRG and time evolution we use ITensor C++ library~\cite{itensor}, where we employ the codes for modified TDVP provided by the authors of~\cite{tdvp2020}.
	For numerical details see Supplemental Materials (SM).
	With this methods we study the case of six fermions (three in each of the components) similarly as in \cite{Syrwid2021a}.
	We assume no intracomponent interactions and probe the stability of the domain by analyzing $\langle x_\text{CM}\rangle$ as in the two-body case.
	Results are presented in Fig.~\ref{fig2}(c).
	The system dynamics for large $|\widetilde g|$ is similar to the two-body problem confirming that the domain can be stabilized over many trap periods due to the strong attraction also in systems involving more than just two particles.
	
	One may suspect that also in the many-body scenario bound states are excluded from the dynamics due to the special properties of the initial state.
	To determine whether this suggestion is true, we deepen our understanding by  investigating many-body solutions of bosonic LL and fermionic YG models.
	Not only do we wish to track down the origins of the domain stabilization visible in numerical results but also ask whether the similarity in the dynamics we found is a universal feature of 1D quantum gases.
	
	\textbf{General repulsive dynamics.}---We employ the LL model of $N$ bosons as a representative of bosonic contact-interacting systems.
	The Hamiltonian reads $H_{\text{LL}}=-\hbar^2/2m\sum_i\partial_{x_i}^2+g\sum_{i<j}\delta(x_i-x_j)$~\cite{Lieb1963}.
	Let us recall that eigenstates of the system take the form of Bethe Ansatz wave functions parametrized by $N$ so-called quasimomenta $k_1, \ldots, k_N$.
	In the regime of strong attraction, eigenstates may be divided into two distinct families~\cite{Durr2009} (see also SM).
	The first one corresponds to states with all real quasimomenta that  for $g \to -\infty$ tend to  $k$'s describing states of the infinitely repulsive system.
	These states reveal anticorrelations characteristic for fermionized system, and in consequence vanish when two particle coordinates are equal, i.e, $x_i=x_{j\neq i}$.
	The remaining states fall into the category of bound states having at least two complex quasimomenta~\cite{Durr2009,Syrwid2021}.
	In analogy to the two-body case, bound states are strongly localized in $N$-dimensional configuration space, vanishing for configurations $\{x_1,\ldots,x_N\}$ where $x_i\neq x_j$ for all $i\neq j$.
	Generally, in the many-body case particles may form $r$-body clusters described by a set of $r$ complex quasimomenta called string solutions \cite{Castin2001,Syrwid2021}.
	In particular, the ground state of the system corresponds to $N$-body cluster given by the McGuire's solution~\cite{Mcguire1964,Thacker1981}.
	
	Consider now a dynamical problem with an arbitrary initial state that has zero probability density of finding two particles at the same position, i.e., described by a  wave function vanishing at $x_i=x_{j\neq i}$.
	Evolution of such a state may be understood via decomposition into $H_{\text{LL}}$ eigenbasis, similarly as in~\eqref{eq::dyn}.
	From the boundlike states' properties, it is clear that for $g\to - \infty$ the only eigenstates participating in the evolution belong to the first family (see also \cite{Girardeau2010a}).
	In this limit even a state consisting of a single complex conjugate pair of quasimomenta is orthogonal to the considered initial state.
	Consequently, the strongly attractive LL system with an initial state vanishing  at $x_i=x_{j\neq i}$ should reveal the repulsive dynamics. 
	
	The same effect can be expected in the presence of an arbitrary confining potential $U(x)$ due to a similar eigenstate structure which for $ g \to -\infty$ belongs to either boundlike or TG-like class~\cite{Volosniev2014}.
	This should not be surprising since a strong attraction induces length scales much shorter than characteristic lengths related to the trapping potential.
	This is why the repulsive dynamics of the strongly attractive system prepared initially in a state vanishing at $x_i=x_{j\neq i}$ seems to be a generic feature of 1D bosonic contact-interacting systems.
	
	Complementary to the bosonic case, we consider a representative of two-component fermionic contact-interacting systems: YG model given by the Hamiltonian $  H_{\text{YG}}=-\hbar^2/2m \sum_{\sigma}\sum_{j=1}^{N_\sigma}\partial_{x_{j}^\sigma}^2 +g\sum_{j=1}^{N_\downarrow} \sum_{s=1}^{N_\uparrow} \delta(x_{j}^\downarrow-x_{s}^\uparrow),$ where $N_{\downarrow,\uparrow}$ denotes number of $\sigma=\downarrow,\uparrow$ fermions~\cite{Yang1967}.
	Similarly to the LL model, the system is Bethe-Ansatz-solvable.
	This time, however, solutions are parameterized with $N=N_\uparrow+N_\downarrow$ quasimomenta and $N_\uparrow$ ($N_\uparrow\leq N_\downarrow$) numbers called spin-roots.
	Again, in the limit $ g \to -\infty$, the eigenstates may be divided into two groups (see SM).
	The first one is characterized by exclusively real quasimomenta and corresponds to states similar to the eigenstates in $ g \to \infty$ limit.
	They vanish when two opposite-spin fermions meet, i.e., $x_{i}^\uparrow=x_{j}^\downarrow$.
	The remaining states, with at least two complex quasimomenta, vanish for configurations $
	\{x_{1}^\downarrow,\ldots,x_{N_\downarrow}^\downarrow,x_{1}^\uparrow,\ldots,x_{N_\uparrow}^\uparrow\}$ such that $x_{i}^\uparrow\neq x_{j}^\downarrow$ for all $i$ and $j\neq i$.
	What distinguishes the fermionic bound states from the bosonic ones is the lack of  strings of quasimomenta longer than $r=2$.
	
	Analogously to the LL model, in case of the initial state being orthogonal to all bound states  (the state vanishing at $x_{i}^\uparrow=x_{j}^\downarrow$), the strongly attractive system will evolve as if the interactions were strongly repulsive.
	To see that the effect survives also in the presence of any external potential $U(x)$, one can follow the same argument as in the bosonic LL case.
	These observations elucidate the domain stability visible in the numerical results, cf. Fig.~\ref{fig2}(c).
	Indeed, in the considered case the initial state is characterized by vanishing probability density when two opposite-spin fermions sit on top of each other, fulfilling necessary conditions for repulsive evolution of strongly attractive systems.
	
	Note that the above-presented criterion for initial states allows for proposal of different setups, where stability despite strong attraction may emerge.
	In Fig.~\ref{fig2}(d) we present an exemplary protocol involving antiferromagnetic chain of two-component fermions trapped in a deep 1D optical lattice.
	Surprisingly, when the lattice potential is suddenly switched off, the presence of a strong contact attraction should lead to a metastable antiferromagnetic order. Let us also note that a similar, repulsive behavior in optical lattice systems was proposed for initially separated, strongly attractive dipolar bosons~\cite{Barbiero2015}.
	
    Finally, let us touch upon the differences between dynamics for large, but finite, positive and negative interactions with the same $|g|$.  In the case of metastable sTG state in the Bose gas, such a relation is known for several observables. For example, the energy of such a state is higher on the attractive side~\cite{Astrakharchik2005}. Similarly, local two-body correlations are enhanced in the sTG case~\cite{Kormos2011}. In the nonequilibrium scenario involving many eigenstates and their overlaps with the initial state, a simple understanding is not at hand. Nevertheless, let us provide some general remarks. For the sTG Bose gas in the low-energy hydrodynamic regime described by the Luttinger theory parameter $K<1$, whereas for the repulsive gas $K>1$ ~\cite{Panfil2013}. Consequently, the dynamics is characterized by phonons propagating slower in the repulsive case. Moreover, in general the system eigenenergies close to $|g| \to \infty$ take the form \cite{Volosniev2014} $E_n=E_n^\text{TG}-\alpha_n/g$, where $E_n^\text{TG}$ is the $n$-th energy level in the TG limit. Typically, $\alpha_n>0$ grows with $n$ (higher states fermionize more slowly with $|g|$ \footnote{We checked that for two harmonically trapped atoms and for the Lieb-Liniger model}). 
    Therefore, the energy differences entering expressions for observables as frequencies governing their evolution are higher on the attractive side than in the TG limit. Conversely,  finite repulsive interactions admit smaller energy separations than in the limit $g\to \infty$. 
    In the specific case of two attractively interacting atoms in harmonic trap, one recovers energy level separation that is larger than the strict TG limit energy spacing given by $2 \hbar \omega$.
    Motion of $\langle x_\text{CM} \rangle$ is mostly governed by a small number of such frequencies corresponding to states with the largest overlaps and this observation explains why $T_\text{mix}(g)>T_\text{mix}(-g)$ for large enough $g>0$ (see Fig.~\ref{fig2}(b)).
	
\begin{figure}[t!]
		\includegraphics[width=\linewidth]{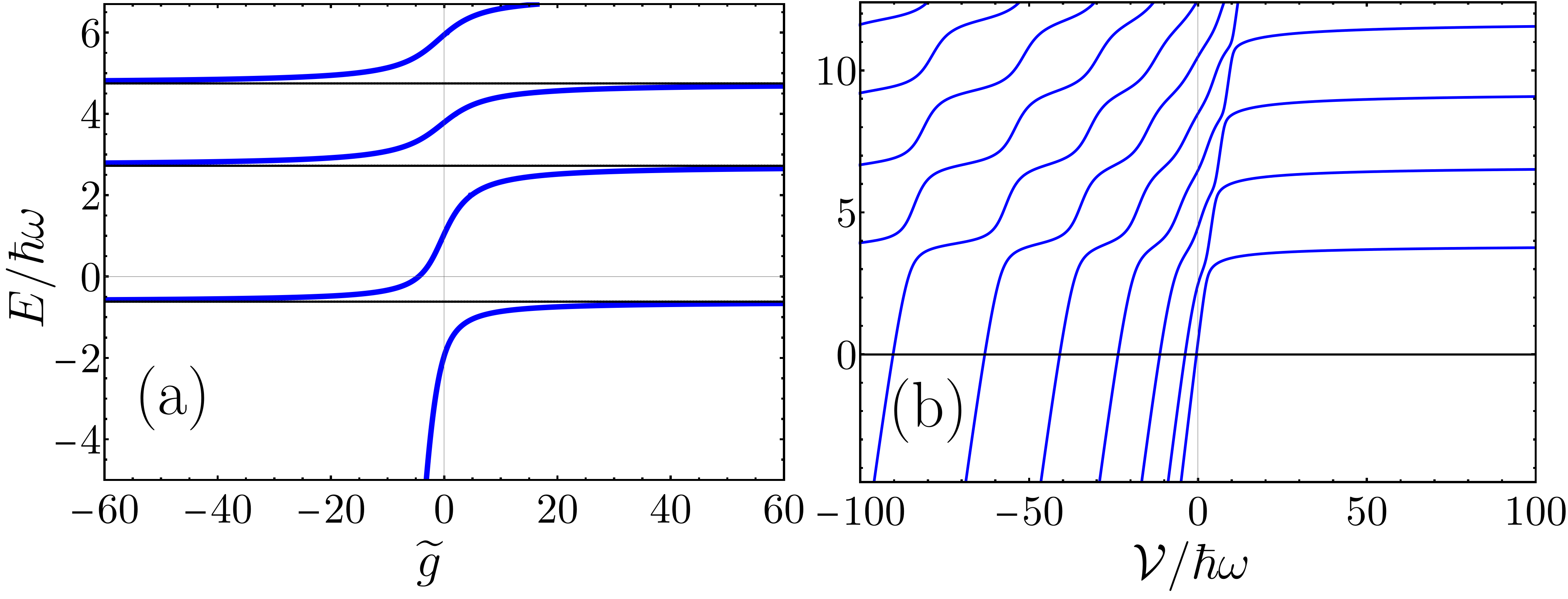}
		\caption{Panel (a): Symmetric state spectrum of the relative Hamiltonian for two particles in a harmonic trap interacting via $V(x)=g\delta (x)+V_\text{well}^a(x)$
			with $a=2d$ and $\mathcal{V}=-2.5\hbar\omega$.
			Note the correspondence  between the $g\to \pm \infty$ limits, similarly to the purely contact interacting case, cf., Fig.~\ref{fig1}(c).
			Panel (b): analogous spectrum but for purely finite-range potential $V(x)=V_\text{well}^a(x)$ with $a=2d$ as a function of well depth $\mathcal{V}$.
			Subsequent states become bound with  decrease of $\mathcal{V}$.
		}\vspace{-0.2cm}
		\label{fig3}
	\end{figure}

	\textbf{Beyond contact interactions.}---It is worthwhile to ask whether some traces of repulsive dynamics can be found in systems with nonlocal interactions.
	One may suspect that an extension of a short-range potential by a finite-range (FR) part, i.e., $V(x)=g \delta(x)+V_\text{FR}(x)$, should not break the above-mentioned similarity between $g \to \pm \infty$. 
	Intuitively, in such a case the strong contact attraction should still lead to states that are either deeply bound or vanishing for $x_i=x_{j\neq i}$.
	We confirm that (see SM) for exactly solvable case of two particles in a harmonic trap interacting via potential $V(x)=g\delta(x)+V_\text{well}^a(x)$, where $V_\text{well}^a(x)= \mathcal{V}[\theta(x+a)-\theta(x-a)]$ is the well potential of width $a$ and depth $ \mathcal{V}$, cf., Fig.~\ref{fig3}(a).
	This observation may be connected to the recent paper \cite{Odziejewski2020} reporting possible existence of bosonic quantum droplets when strong contact repulsion is supplemented by a weak dipolar attraction. 
	In contrast to commonly studied quantum-fluctuations-stabilized droplets, in this case the Lee-Huang-Yang correction does not play a decisive role and sufficiently strong repulsive interactions are the key ingredient of their stability.
	The corresponding droplet wave function should vanish at $x_i=x_{j\neq i}$.
	Similarly to the sTG gas~\cite{Astrakharchik2005,Chen2010}, 
	the droplet state should not collapse after a interaction quench to strong contact attraction despite both short- and finite-range attractive interactions, see Fig.~\ref{fig2}(e).
	
	The picture is very different when we consider strong and purely finite-range interactions.
	Fig.~\ref{fig3}(b) presents energies of symmetric states determined for two particles in a harmonic trap interacting via $V(x)=V_\text{well}^a(x)$.
	Contrary to the case of pointlike interactions, finite-range potentials may exhibit more than one bound state, depending on the potential parameters $ \mathcal{V}$ and $a$.
	As we tune $ \mathcal{V}$ towards infinitely strong attraction while keeping $a$ fixed, more and more states become bound.
	Nevertheless, in the case of exclusively finite-range interactions, the spatial width of even deeply bound states remains finite and is characterized by the potential range $a$.
	Therefore, in the context of the repulsive dynamics, if the highest bound state is significantly deep, the orthogonality of the initial state to the bound states requires much stronger separation between attractively interacting particles, i.e. on a range at least comparable with $a$.
	Let us note, however, that in the specific case of three-delta potential, the repulsive dynamics was recently reported~\cite{Syrwid2021a}. 
	
	\textbf{Conclusions.}---We have found that strongly attractive short-range interacting gases in 1D may exhibit identical dynamical behavior as the strongly repulsive ones.
	Such a phenomenon takes place when the system is initially prepared in a proper state with well separated interacting particles, which can be realized experimentally via specific interaction or confinement quenches.
	We provide arguments supporting the wide universality of the mechanism.
	Both the strongly repulsive and strongly attractive regimes, $g \to \pm \infty$,  are attainable by means of Feshbach resonances.
	In this context, the discussed interaction quench in dipolar quantum droplets and the dynamics of the domain wall in a two-component Fermi gas seem particularly interesting and experimentally feasible.

		\textbf{Acknowledgments.}---The authors would like to thank Artem Volosniev, Tomasz Sowiński and Vicky C. Arizona for fruitful discussions.
		M. \L{}. acknowledges the support from the (Polish) National Science Center Grant 2018/29/B/ST2/01308.
		A. S. acknowledges the support from Olle Engkvists stiftelse.
		P. T. G. is financed from the (Polish) National Science Center Grants 2018/31/N/ST2/01429  and 2020/36/T/ST2/00065. 
		K. Rz. is supported from the (Polish) National Science Center Grant 2018/29/B/ST2/01308.
		Center for Theoretical Physics of the Polish Academy of Sciences is a member of the National Laboratory of Atomic, Molecular and Optical Physics (KL FAMO).

	\bibliography{library}
	\newpage
	

\section*{Supplementary material}

# Repulsive dynamics of strongly attractive one-dimensional quantum gases: Supplemental materials

Maciej Łebek[ID],[1,2,*] Andrzej Syrwid[ID],[3,†] Piotr T. Grochowski[ID],[1,4,‡] and Kazimierz Rzążewski[ID][1,§]

[1] *Center for Theoretical Physics, Polish Academy of Sciences, Aleja Lotników 32/46, 02-668 Warsaw, Poland*
[2] *Institute of Theoretical Physics, University of Warsaw, Pasteura 5, 02-093 Warsaw, Poland*
[3] *Department of Physics, The Royal Institute of Technology, Stockholm SE-10691, Sweden*
[4] *ICFO - Institut de Ciències Fotòniques, The Barcelona Institute of Science and Technology, Av. Carl Friedrich Gauss 3, 08860 Castelldefels (Barcelona), Spain*
(Dated: July 12, 2021)

## LIEB-LINIGER AND YANG-GAUDIN MODELS IN THE REGIME OF STRONG ATTRACTION

First, we consider the Lieb-Liniger model of $N$ bosons on a ring with length $L$ described by the Hamiltonian [1–3]

$$H_{LL} = -\sum_i \partial_{x_i}^2 + 2c \sum_{i<j} \delta(x_i - x_j). \tag{1}$$

We focus on the regime of strongly attractive interactions, i.e., $cL/N$ large and negative. Eigenstates defined in the fundamental domain $D = \{(x_1, \ldots, x_N) : x_1 \leq \ldots \leq x_N\}$ take the form [1, 3]

$$\psi_{\{k\}}(\{x\}) \propto \sum_{\pi \in S_N} \text{sgn}(\pi) \prod_{j>s} \left(k_{\pi(j)} - k_{\pi(s)} - ic\right) \exp\left(i \sum_n k_{\pi(n)} x_n\right), \tag{2}$$

and are parametrized by $N$ numbers called quasimomenta $k_1, \ldots, k_N$ satisfying the Bethe equations [1, 3, 4]

$$e^{ik_j L} = -\prod_{s=1}^N \frac{k_j - k_s + ic}{k_j - k_s - ic}, \qquad j = 1, 2, \ldots, N. \tag{3}$$

Let us consider the case of states from the first family i.e. parameterized by purely real quasimomenta. In such a case (3) may be cast into the following form [3]

$$k_j L + 2 \sum_{s=1}^N \arctan\left(\frac{k_j - k_s}{c}\right) = 2\pi I_j^p, \tag{4}$$

where $I_j^p$ are arbitrary integers (half-integers) when $N$ is odd (even). For large $|c|$ we may expand arctan getting

$$k_j = \frac{2\pi}{L} I_j^p - \frac{2}{cL} \sum_{s=1}^N (k_j - k_s). \tag{5}$$

Note that for $c \to \pm\infty$ the quasimomenta from the first family tend to the same values $k_j = 2\pi I_j^p/L$ independently of the sign of $c$. The corresponding wave functions vanish when two particles approach each other, i.e., for configurations $\{x_1, \ldots, x_N\}$ such that $x_i = x_j$ for at least one pair $i \neq j$.

Now we turn to the states belonging to the second family, characterized by at least two complex quasimomenta. In general, the particles may form $r$-body clusters described by a set of $r$ quasimomenta [3, 5]

$$k_\alpha = k_0 + i\frac{c}{2}(r + 1 - 2\alpha), \tag{6}$$

where $\alpha = 1, 2, \ldots, r$ and $k_0 \in \mathbb{R}$ is the momentum of the cluster. We wish to study now the properties of wave functions corresponding to such bound states. Let us initially focus on the case of just one two-body cluster represented by a single complex conjugate pair of quasimomenta and $N - 2$ remaining real quasimomenta, i.e., $k_1 = i\frac{c}{2}, k_2 = -i\frac{c}{2}$ and $k_3, k_4, \ldots, k_N \in \mathbb{R}$. The corresponding wave function in the fundamental domain may be written as follows

$$\psi_{\{k\}}(\{x\}) \propto \sum_{\pi \in S_N} \text{sgn}(\pi) \prod_{j>s} \left(k_{\pi(j)} - k_{\pi(s)} - ic\right) \exp\left(i \sum_{\substack{n \\ \pi(n) \neq 1,2}} k_{\pi(n)} x_n\right) f(x_{\pi^{-1}(1)}, x_{\pi^{-1}(2)}), \tag{7}$$

with $f(x,y) = \theta(x-y)\exp\left[-|c|(x-y)/2\right]$ where $\theta(x)$ denotes Heaviside step function. Note that $f(x,y)$ vanishes rapidly far from $x-y \approx 0_+$. Therefore, for a very large $|c|$ (remember that here $c < 0$), each term in the summation in Eq. (7) is exponentially localized at $x_{\pi^{-1}(1)} - x_{\pi^{-1}(2)} \to 0_+$. Due to this $\Psi \to 0$ for any configuration $\{x_1, x_2, \ldots, x_N\}$ in which $x_i \neq x_j$ for all $j \neq i$. In consequence, state (7) is orthogonal to any state $\Psi_0$ vanishing for configurations where at least two particles' positions coincide, due to the complete mismatch of supports.

Similar calculation may be performed for states with clusters involving more particles. The conclusion is the same: in the limit $c \to -\infty$ all these states are orthogonal to $\Psi_0$ as opposed to the states from the first family that generally have a nonzero overlap with $\Psi_0$. Finally, let us observe that the wave function becomes more and more localized in the configuration space when dealing with clusters involving more and more particles (see also discussion in [6]). The ground state, corresponding to the $N$-body cluster takes the form [7]

$$\psi_{\text{GS}}(x_1, \ldots, x_N) \propto \exp\left(-\frac{|c|}{2}\sum_{j>s}|x_j - x_s|\right). \tag{8}$$

In a specific example of $N = 3$ system with a one bound pair, we see that such a state is localized on two-dimensional planes $x_1 = x_2$, $x_2 = x_3$ and $x_1 = x_3$ and has larger support (in the sense of dimension) than the McGuire's ground state (8) localized on a line given by the condition $x_1 = x_2 = x_3$.

Similar structure can be also observed in the Yang-Gaudin model [8] describing two-component Fermi gas on a ring with circumference $L$. The Hamiltonian reads

$$H = -\left(\sum_{j=1}^{N_\downarrow}\partial^2_{x_j^\downarrow} + \sum_{j=1}^{N_\uparrow}\partial^2_{x_j^\uparrow}\right) + 2c\sum_{j=1}^{N_\downarrow}\sum_{s=1}^{N_\uparrow}\delta(x_j^\downarrow - x_s^\uparrow), \tag{9}$$

where $N_\downarrow(N_\uparrow)$ denotes number of $\downarrow$ ($\uparrow$) fermions. The model again can be solved via the Bethe Ansatz technique. The solutions are parameterized with $N = N_\uparrow + N_\downarrow$ quasimomenta $k_1, \ldots, k_N$ and $N_\uparrow$ numbers $\Lambda_1 \ldots, \Lambda_{N_\uparrow}$ called spin-roots (without loss of generality we assume $N_\downarrow \geq N_\uparrow$). The quasimomenta and spin-roots have to satisfy the following Bethe Ansatz equations [3]

$$e^{ik_j L} = \prod_{n=1}^{N_\uparrow}\frac{k_j - \Lambda_n + i\frac{c}{2}}{k_j - \Lambda_n - i\frac{c}{2}}, \qquad j = 1, \ldots, N, \tag{10}$$

$$\prod_{j=1}^{N}\frac{\Lambda_m - k_j + i\frac{c}{2}}{\Lambda_m - k_j - i\frac{c}{2}} = \prod_{\substack{n=1 \\ n\neq m}}^{N_\uparrow}\frac{\Lambda_m - \Lambda_n + ic}{\Lambda_m - \Lambda_n - ic}, \qquad m = 1, \ldots, N_\uparrow. \tag{11}$$

The wave functions can be cast into the determinant form [3, 9, 10]

$$\psi(\{x^\downarrow\}_{N_\downarrow}, \{x^\uparrow\}_{N_\uparrow}, \{k\}_N, \{\Lambda\}_{N_\uparrow}) \propto \sum_{\pi \in S_{N_\uparrow}} \text{sgn}\pi\, \mathcal{W}_\pi^\uparrow \det\Phi_{N\times N}, \tag{12}$$

with

$$\mathcal{W}_\pi^\uparrow = \prod_{j<s}^{N_\uparrow}[i(\Lambda_{\pi(j)} - \Lambda_{\pi(s)}) + c\,\text{sign}(x_s^\uparrow - x_j^\uparrow)], \tag{13}$$

and the $N \times N$ matrix $\Phi_{N\times N}$ constructed from two matrices of size $N \times N_\downarrow$ and $N \times N_\uparrow$ involving plane waves

$$\Phi_{N\times N} = \left(\left[\prod_{s=1}^{N_\uparrow}\mathcal{A}_j(\Lambda_{\pi(s)}, x_l^\downarrow - x_s^\uparrow)e^{ik_j x_l^\downarrow}\right]\Big|\left[\prod_{s\neq m}^{N_\uparrow}\mathcal{A}_j(\Lambda_{\pi(s)}, x_m^\uparrow - x_s^\uparrow)e^{ik_j x_m^\uparrow}\right]\right)_{\substack{j=1,\ldots,N \\ l=1,\ldots,N_\downarrow \\ m=1,\ldots,N_\uparrow}}, \tag{14}$$

where

$$\mathcal{A}_j(\Lambda, x) = i(k_j - \Lambda) + \frac{c}{2}\text{sign}(x). \tag{15}$$

As in the Lieb-Liniger model the eigenstates can be divided into two groups. The first one is characterized by exclusively real quasimomenta. In such a case Eqs. (10) and (11) can be rewritten as follows [3, 11, 12]

$$k_j L + 2\sum_{n=1}^{N^\uparrow} \arctan\left[\frac{2(k_j - \Lambda_n)}{c}\right] = 2\pi \mathcal{Y}_j^p, \quad 2\sum_{j=1}^{N} \arctan\left[\frac{2(\Lambda_m - k_j)}{c}\right] - 2\sum_{\substack{n=1 \\ n\neq m}}^{N_\uparrow} \arctan\left[\frac{2(\Lambda_m - \Lambda_n)}{c}\right] = 2\pi J_m^p, \quad (16)$$

with $\mathcal{Y}_j^p (J_m^p)$ being integers when $N_\uparrow (N_\downarrow)$ is even and half-odd integers when $N_\uparrow (N_\downarrow)$ is odd (even). For large c we may expand in $c^{-1}$ getting

$$k_j \approx \frac{2\pi}{L}\mathcal{Y}_j^p\left(1 - \frac{4N^\uparrow}{cL}\right), \quad \Lambda_m = \gamma_m c, \quad (17)$$

where constants $\gamma_m$ can be determined as in [11]. In the limit $c \to -\infty$ real quasimomenta tend to the quasimomenta of $c \to \infty$ system, $k_j = \frac{2\pi}{L}\mathcal{Y}_j^p$, whereas spin-roots differ from the $c \to \infty$ case with sign. However, that difference does not have any impact on the form of the wave function and states described by (17) in the limit $c \to -\infty$ vanish when $x_i^\uparrow = x_j^\downarrow$, due to the fermionization between components.

The second group consists of bound states with at least two complex momenta. Contrary to the case of bosons, in the $c \to -\infty$ limit one can only have strings with length 2, describing bound pairs of fermions from different components

$$k_{j,\pm} = \Lambda_j \pm i\frac{c}{2}. \quad (18)$$

For example the ground state consists of $N_\uparrow$ such pairs. As in the case of the Lieb-Liniger model, an appearence of quasimomenta with imaginary parts $\propto c$ leads to exponential localization (with a decay rate proportional to $|c|$) of the Bethe Ansatz wave function (12). Bound states vanish for configurations $\{x_1^\downarrow, \ldots, x_{N_\downarrow}^\downarrow, x_1^\uparrow, \ldots, x_{N_\uparrow}^\uparrow\}$ such that $x_i^\uparrow \neq x_j^\downarrow$ for all $i$ and $j$. Initial states vanishing for any configuration $\{x_1^\downarrow, \ldots, x_{N_\downarrow}^\downarrow, x_1^\uparrow, \ldots, x_{N_\uparrow}^\uparrow\}$ with $x_i^\uparrow = x_j^\downarrow$ for some $i, j$ will be orthogonal to bound states, due to complete mismatch of supports.

## TWO ATOMS IN HARMONIC TRAP—SOLUTION

We consider two atoms of mass $m$ in a harmonic trap with frequency $\omega$. The atoms interact via potential consisting of a short-range part given by delta function and non-local interaction modeled with a well potential characterized by range (width) $a$. The Hamiltonian reads

$$H = -\frac{\hbar^2}{2m}\left(\partial_x^2 + \partial_y^2\right) + \frac{1}{2}m\omega^2(x^2 + y^2) + V(x - y) \quad (19)$$

with $V(x-y) = g\delta(x-y) + V_{\text{well}}^a(x-y)$, where $V_{\text{well}}^a(x) = V[\theta(x+a) - \theta(x-a)]$. The solution of the problem closely follows earlier works on similar problems involving delta and well potentials [4, 13–15]. We introduce new variables $R = (x+y)/\sqrt{2}$ and $r = (x-y)/\sqrt{2}$ bringing our Hamiltonian to the form $H = H_{\text{CM}} + H_{\text{rel}}$, where $H_{\text{CM}}$ is the harmonic oscillator Hamiltonian and $H_{\text{rel}} = -\frac{\hbar^2}{2m}\partial_r^2 + \frac{m\omega^2}{2}r^2 + V'(r)$. Here, $V'(r) = g'\delta(r) + V_{\text{well}}^{a'}(r)$ with $g' = g/\sqrt{2}$ and $a' = a/\sqrt{2}$. We focus on the eigenstates of $H_{\text{rel}}$. The solutions are either symmetric (+) or antisymmetric (−) functions of $r$ and read

$$\psi_+(r) = \begin{cases} A_+ \tilde{U}_{\frac{\sigma}{2},\frac{1}{2}}(r) & |r| \geq a' \\ B_+ \tilde{F}_{\frac{\sigma+\bar{V}}{2},\frac{1}{2}}(r) + C_+ \tilde{U}_{\frac{\sigma+\bar{V}}{2},\frac{1}{2}}(r) & |r| < a' \end{cases}, \quad (20)$$

$$\psi_-(r) = \begin{cases} A_- \frac{r}{d} \tilde{U}_{\frac{\kappa}{2},\frac{3}{2}}(r) & |r| \geq a' \\ B_- \frac{r}{d} \tilde{F}_{\frac{\kappa+\bar{V}}{2},\frac{3}{2}}(r) & |r| < a' \end{cases}, \quad (21)$$

where $d = \sqrt{\hbar/m\omega}$ is the characteristic oscillator length, $\bar{V} = V/\hbar\omega$, $\kappa = 3/2 - E/\hbar\omega$, and $\sigma = 1/2 - E/\hbar\omega$. For convenience, we introduced functions $\tilde{F}_{\xi,\zeta}(r) = \exp\left(-r^2/2d^2\right){}_1F_1(\xi;\zeta;r^2/d^2)$ and $\tilde{U}_{\xi,\zeta}(r) = \exp\left(-r^2/2d^2\right)U(\xi,\zeta,r^2/d^2)$, where $U$ and ${}_1F_1$ are confluent hypergeometric functions of the first and second type, respectively. Energies of the system and relations between coefficients $A_\pm, B_\pm$ can be determined by imposing continuity and $\delta$-potential conditions

at the points $r = 0, \pm a'$. We start with the symmetric solutions, where the continuity of the wave function at $r = \pm a'$ yields

$$A_+ U\left(\frac{\sigma}{2}, \frac{1}{2}, \frac{a'^2}{d^2}\right) = B_{+1}F_1\left(\frac{\sigma + \bar{V}}{2}; \frac{1}{2}; \frac{a'^2}{d^2}\right) + C_+ U\left(\frac{\sigma + \bar{V}}{2}, \frac{1}{2}, \frac{a'^2}{d^2}\right), \tag{22}$$

whereas the continuity of the derivative at $r = \pm a'$ leads to

$$-A_+\sigma U\left(1+\frac{\sigma}{2}, \frac{3}{2}, \frac{m\omega}{\hbar}a'^2\right) = 2B_+(\sigma+\bar{V})_1F_1\left(1+\frac{\sigma+\bar{V}}{2}; \frac{3}{2}; \frac{m\omega}{\hbar}a'^2\right) - C_+(\sigma+\bar{V})U\left(1+\frac{\sigma+\bar{V}}{2}, \frac{3}{2}, \frac{m\omega}{\hbar}a'^2\right). \tag{23}$$

Finally, the $\delta$-potential condition at $r = 0$ gives [15]

$$C_+\hbar\omega d\frac{\sqrt{\pi}(\sigma+\bar{V})}{\Gamma\left(1+\frac{\sigma+\bar{V}}{2}\right)} + g'\left[B_+ + C_+\frac{\sqrt{\pi}}{\Gamma\left(\frac{1}{2}+\frac{\sigma+\bar{V}}{2}\right)}\right] = 0. \tag{24}$$

This leads to the following relations

$$B_+ = -\Xi_{\sigma+\bar{V}}C_+, \tag{25}$$

$$A_+ = \frac{\sigma+\bar{V}}{\sigma}\frac{2\Xi_{\sigma+\bar{V}\,1}F_1\left(1+\frac{\sigma+\bar{V}}{2}; \frac{3}{2}; \frac{a'^2}{d^2}\right) + U\left(1+\frac{\sigma+\bar{V}}{2}, \frac{3}{2}, \frac{a'^2}{d^2}\right)}{U\left(1+\frac{\sigma}{2}, \frac{3}{2}, \frac{a'^2}{d^2}\right)}C_+, \tag{26}$$

with $\Xi_\alpha = \sqrt{\pi}\left[\hbar\omega d\alpha/g'\Gamma\left(1+\frac{\alpha}{2}\right) + 1/\Gamma\left(\frac{1}{2}+\frac{\alpha}{2}\right)\right]$, and by the continuity condition one finds

$$\frac{2\Xi_{\sigma+\bar{V}\,1}F_1\left(1+\frac{\sigma+\bar{V}}{2}; \frac{3}{2}; \frac{a'^2}{d^2}\right) + U\left(1+\frac{\sigma+\bar{V}}{2}, \frac{3}{2}, \frac{a'^2}{d^2}\right)}{U\left(\frac{\sigma+\bar{V}}{2}, \frac{1}{2}, \frac{a'^2}{d^2}\right) - \Xi_{\sigma+\bar{V}\,1}F_1\left(\frac{\sigma+\bar{V}}{2}; \frac{3}{2}; \frac{a'^2}{d^2}\right)} = \frac{\sigma}{\sigma+\bar{V}}\frac{U\left(1+\frac{\sigma}{2}, \frac{3}{2}, \frac{a'^2}{d^2}\right)}{U\left(\frac{\sigma}{2}, \frac{1}{2}, \frac{a'^2}{d^2}\right)}, \tag{27}$$

from which we determine energies of symmetric states. For antisymmetric solutions the continuity at $r = \pm a'$ implies

$$A_-U\left(\frac{\kappa}{2}, \frac{3}{2}, \frac{a'^2}{d^2}\right) = B_{-1}F_1\left(\frac{\kappa+\bar{V}}{2}; \frac{3}{2}; \frac{a'^2}{d^2}\right), \tag{28}$$

while the continuity of the derivative at $r = \pm a'$ gives

$$-A_-\kappa U\left(1+\frac{\kappa}{2}, \frac{5}{2}, \frac{a'^2}{d^2}\right) = \frac{2}{3}(\kappa+\bar{V})B_{-1}F_1\left(1+\frac{\kappa+\bar{V}}{2}; \frac{5}{2}; \frac{a'^2}{d^2}\right). \tag{29}$$

Eventually, we arrive at the equation for energies of antisymmetric states

$$2(\kappa-V')_1F_1\left(1+\frac{\kappa-V'}{2}, \frac{5}{2}, \frac{m\omega}{\hbar}a'^2\right)U\left(\frac{\kappa}{2}, \frac{3}{2}, \frac{m\omega}{\hbar}a'^2\right) + 3\kappa_1F_1\left(\frac{\kappa-V'}{2}, \frac{3}{2}, \frac{m\omega}{\hbar}a'^2\right)U\left(1+\frac{\kappa}{2}, \frac{5}{2}, \frac{m\omega}{\hbar}a'^2\right). \tag{30}$$

The wave functions (20) are determined from relations between coefficients $A_\pm, B_\pm, C_+$ and a normalization condition. As a result the two-body wave functions are of the form

$$\Psi_{n,\mu}(x,y) = \phi_n\left(\frac{x+y}{\sqrt{2}}\right)\psi_\mu\left(\frac{x-y}{\sqrt{2}}\right), \tag{31}$$

where $\phi_n$ are eigenstates of harmonic oscillator and $\mu$ denotes either $\sigma$ or $\kappa$, depending on the symmetry of the state. The spectrum for the case of two bosons (fermions) is determined by taking only states with $n$ and $\mu$ such that the whole function has a proper symmetry.

## DETAILS OF NUMERICAL SIMULATIONS

The many-body dynamics of the system involving 6 particles is studied by employing matrix product states (MPS) [16, 17] Ansatz with open boundary conditions (OBC). The system size is chosen to be equal $L = 10$ having the hard-wall boundaries at $x_{\text{OBC}} = \pm L/2$. The space is discretized into $N_s = 50$ sites located at $x_{j=1,\ldots,N_s} = -L/2 + \Delta(j-1/2)$ where $\Delta = L/N_s = 0.2$. In addition to OBC we have harmonic trap centered at $x = 0$ with the

frequency $\omega = \sqrt{2}$. This choice guarantees that particles cannot explore regions close to the hard-wall boundaries and thus do not feel OBC. To prepare the initially separated state we employ a standard density matrix renormalization group (DMRG) method [16–20] in the presence of a Gaussian barrier $V_b(x) = A e^{-x^2/2\sigma^2}$ with $A = 40$ and $\sigma = 0.2$. The algorithm is initialized with a state where fermions belonging to different spin components occupy sites in different halves of the harmonic trap, far from $x = 0$.

In general, time evolution within the MPS approach is computationally demanding and error-prone. Indeed, due to a ballistic increase of an entanglement entropy one has to deal with increasing bond dimension. Partial overcome of this issue came with a recent development of the the time-dependent variational principle (TDVP) algorithm [21–24], within which the system temporal behavior can be studied when keeping a fixed bond dimension (one-site TDVP). Although, such a technique is much less error-prone comparing to earlier strategies like the time evolving block decimation (TEBD) method [25], the bond dimension has to be still first enlarged enough to obtain correct results. This is realized for example as in the "hybrid" TDVP routine, where the initial application of the two-site TDVP scheme allows to reach the desired bond dimension [25–27]. In this work we investigate the dynamics with the help of a very recent approach meticulously described in Ref. [28], where the basis for MPS $|\Psi\rangle$ is enlarged through a subspace expansion by global Krylov vectors $\{|\Psi\rangle, \hat{H}|\Psi\rangle, \ldots, \hat{H}^{k-1}|\Psi\rangle\}$, where $\hat{H}$ is the time evolved system Hamiltonian and $k$ denotes the so-called Krylov order. As it is pointed out in in Ref. [28], this method turns out to be more accurate and more efficient than the two-site TDVP strategy. Additionally, it also gives more reliable results when dealing with beyond-nearest-neighbor interactions. We use the algorithm implementation based on the ITensor library provided by the authors of [28] under the ITensor/TDVP repository.

The numerical simulations we performed with the time step $\Delta t = 0.01$ and $k = 3$. The truncation error of each application of $\hat{H}$ to $|\Psi\rangle$ and the truncation error controlling diagonalization of the sum of the reduced density matrices were set to $10^{-8}$. Unitarity is preserved by imposing no truncation during one-site TDVP sweeps. The global basis expansion is switched off when the bond dimension $\chi$ of the time-evolved MPS reaches the pre-determined maximal value $\chi_{\max} = 750$. In Fig. 1 we present the convergence of our results. Left panels (a)–(d) show comparison between the temporal behavior of $\langle x_{\mathrm{CM}} \rangle$ in $N_s = 50$ sites system consisting of 3 fermions in each spin component for $\widetilde{g} = \pm 100, \pm 300$, determined with different maximal bond dimension $\chi_{\max} = 250, 500, 750$. Right panels (e)–(h) illustrate comparison between the von Neumann entanglement entropies dynamics, $S_{j=25}$, calculated across the central bond (between the central $j = 25$ and $j + 1 = 26$ sites) with different $\chi_{\max} = 250, 500, 750$.

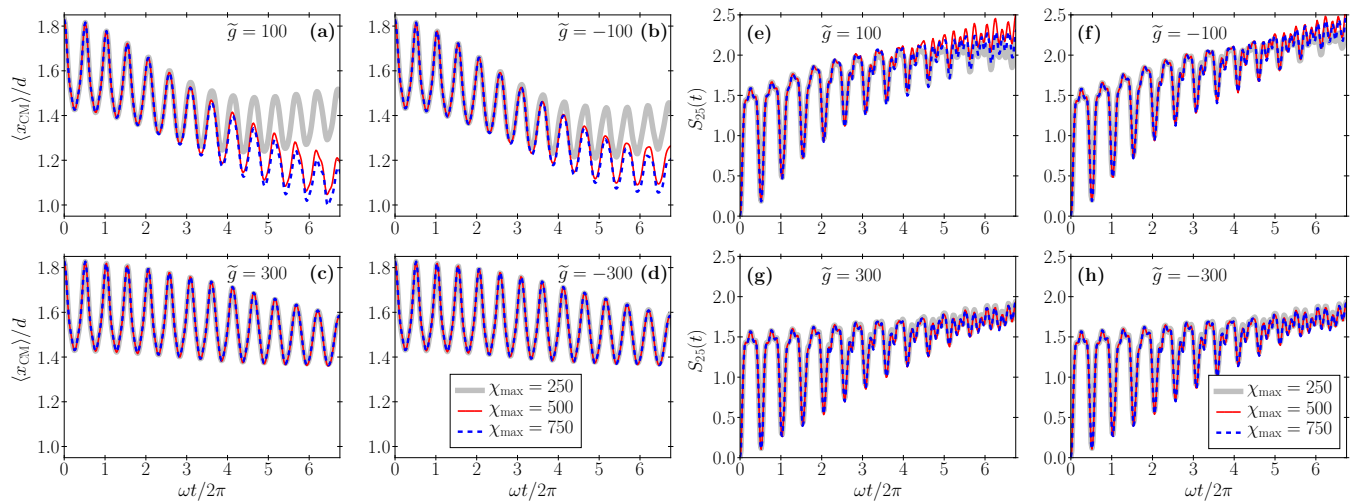

FIG. 1. Left panels (a)–(b): Dynamics of the center of mass position $\langle x_{\mathrm{CM}} \rangle$ of one of the Fermi clouds for different contact interactions strength $\widetilde{g} = \pm 100, \pm 300$ and different maximal bond dimension $\chi_{\max} = 250, 500, 750$. Right panels (e)–(h): The von Neumann entanglement entropy calculated across the central bond between 25th and 26th sites, obtained for different $\widetilde{g} = \pm 100, \pm 300$ and $\chi_{\max} = 250, 500, 750$.

* mlebek@cft.edu.pl
† syrwid@kth.se

‡ piotr@cft.edu.pl
§ kazik@cft.edu.pl


\end{document}